\documentclass[showpacs,10pt,twocolumn,prb]{revtex4-1}
\usepackage{amsmath}
\usepackage{amssymb}
\usepackage{graphics}
\usepackage{epsfig}

\begin{document}


\title{Structure and physical properties of new layered iron oxychalcogenide BaFe$_{2}$OSe$_{2}$}
\author{Hechang Lei,$^{1}$, Hyejin Ryu,$^{1,2}$ John Warren,$^{3}$ A. I. Frenkel,$^{4}$ V. Ivanovski,$^{5}$ B. Cekic$^{5}$ and C. Petrovic$^{1,2}$}
\affiliation{$^{1}$Condensed Matter Physics and Materials Science Department, Brookhaven
National Laboratory, Upton, New York 11973, USA\\
$^{2}$Department of Physics and Astronomy, Stony Brook University, Stony Brook, New York 11794-3800, USA\\
$^{3}$Instrumentation Division, Brookhaven National Laboratory, Upton, New
York 11973, USA\\
$^{4}$Physics Department, Yeshiva University, 245 Lexington Avenue, New York, New York 10016, USA\\
$^{5}$Institute of Nuclear Sciences Vinca, University of Belgrade, Belgrade 11001, Serbia}

\date{\today}

\begin{abstract}
We have successfully synthesized a new layered iron oxychalcogenide BaFe$_{2}$OSe$_{2}$ single crystal. This compound is built up of Ba and Fe-Se(O) layers alternatively stacked along the c-axis. The Fe-Se(O) layers contain double chains of edge-shared Fe-Se(O) tetrahedra that propagate along the b-axis and are bridged by oxygen along the a-axis. Physical property measurements indicate that BaFe$_{2}$OSe$_{2}$ is a semiconductor without the Curie-Weiss behavior up to 350 K. There is a possible long range antiferromagnetic (AFM) transition at 240 K, corresponding to the peak in specific heat measurement and two glassy transitions at 115 K and 43 K. The magnetic entropy up to 300 K is much smaller than the expected value for Fe$^{2+}$ in tetrahedral crystal fields and M\"{o}sbauer spectrum indicates that long range magnetic order is unlikely at 294 K. Both results suggest that a short range magnetic correlations exist above the room temperature.
\end{abstract}

\pacs{75.50.Ee, 75.10.Pq, 75.30.Gw, 75.47.Np}
\maketitle

\section{Introduction}

The discovery of layered iron oxypnictides (Ln(O,F)FeAs, Ln = rare earth
elements, 1111-type) superconductors with $T_{c}$ up to 56\ K,\cite{Kamihara}
has stimulated great interest in mixed-anion materials. Mixed anions from
the same row of the Mendeleev periodic table tend to randomly occupy the
same crystallographic site (anion disorder) because of the relatively
similar sizes. On the other hand, if the anions are from different rows
(oxysulfide or oxyselenide compounds for example), the distinctive
difference of anion sizes and ionic polarization might lead to ordered
occupancy in the different crystallographic sites (anion order), often
forming layered crystal structure.\cite{Clarke}

The mixed-anion compounds have attracted some interest during the
exploration of novel cuprate high temperature superconductors.
Halooxocuprates are example where copper ions are coordinated with four
oxygen ions, forming the CuO$_{2}$ sheet, whereas the halogen ion usually
occupies a so-called apical site.\cite{Adachi} With proper electron or hole
doping, halooxocuprates will become superconductors.\cite{Adachi} Besides
superconductivity, mixed-anion materials also exhibit diverse physical
properties. Copper oxychalcogenides LnCuOCh (Ch = S, Se,\ and Te),
isostructural to 1111-type iron-based superconductors, are wide-gap p-type
semiconductors with transparent p-type conductivity, photoluminescence, and
large third order optical nonlinearity.\cite{Ueda}$^{-}$\cite{Kamioka}
Transition metal oxychalcogenides Ln$_{2}$O$_{2}$TM$_{2} $OCh$_{2}$ (TM =
Mn, Fe, and Co) contain the layers built up by the edge-shared octahedral
unit [TM$_{2}$OCh$_{2}$]$^{2-}$. These materials show strong
electron-electron interactions (Mott insulators) on the two dimensional (2D)
frustrated antiferromagnetic (AFM) checkerboard spin-lattice.\cite{Mayer}$%
^{-}$\cite{Ni}

The materials with [TM$_{2}$OCh$_{2}$]$^{2-}$ layers exhibit similar
structural diversity to iron-based superconductors. This is understandable
since new compounds can be obtained by simply replacing [FeAs]$^{-}$ layers
with [TM$_{2}$OCh$_{2}$]$^{2-}$ layers, such as LnOFeAs $\rightarrow$ Ln$%
_{2} $O$_{2}$TM$_{2}$OCh$_{2}$, AEFFeAs (AE = alkali earth metals) $%
\rightarrow$ AE$_{2}$F$_{2}$TM$_{2}$OCh$_{2}$,\cite{Han}$^{-}$\cite{Liu} and
AFeAs (A = alkali metals) $\rightarrow$ Na$_{2}$Fe$_{2}$OSe$_{2}$.\cite{Wang
XC}$^{-}$\cite{He} Thus, it is of considerable interest to explore the
structural derivatives of AEFe$_{2}$As$_{2}$ among the oxychalcogenide
compounds.

Here, we report the detailed synthesis and physical properties of a new
layered iron oxychalcogenide BaFe$_{2}$OSe$_{2}$ single crystal. Even
though it has the same chemical formula and [TM$_{2}$OCh$_{2}$]$^{2-}$
layers, the structure is different from other iron oxychalcogenides with [TM$%
_{2}$OCh$_{2}$]$^{2-}$ layers. To the best of our knowledge, BaFe$_{2}$OSe$%
_{2}$ is the first layered iron oxychalcogenide with alkali earth metal. It
shows a semiconducting behavior with possible successive spin-glass
transitions at low temperature and short range antiferromagnetic
order above the room temperature.

\section{Experiment}

Single crystals of BaFe$_{2}$OSe$_{2}$ were synthesized by a self-flux
method. Ba rod, Fe powder, Fe$_{2}$O$_{3}$ powder and Se shot were used as
starting materials. BaSe was prereacted by reacting Ba piece with Se shot at
800 ${{}^{\circ }}C$ for 12 hours. BaSe was mixed with other reagents and
intimately ground together using an agate pestle and mortar. The ground
powder was pressed into pellets, loaded in an alumina crucible and then
sealed in quartz tubes with Ar under the pressure of 0.2 atmosphere. The
quartz tubes were heated up to 600 ${{}^{\circ }}C$ in 10 h, kept at 600 ${%
{}^{\circ }}C$ for 12 h, ramped again to 1100 ${{}^{\circ }}C$ in 12 h, kept
at 1100 ${{}^{\circ }}C$ for 12 h, and then cooled slowly at a rate of 3 ${%
{}^{\circ }}C$/h to 800 ${{}^{\circ }}C$, finally the furnace was shut down
and the sample was cooled down to room temperature naturally. Black
plate-like crystals with typical size 2$\times $2$\times $0.5 mm$^{3}$ can
be grown. Except for heat-treatment, all of processes for sample preparation
were performed in glove boxes filled with argon.

The crystal structure of BaFe$_{2}$OSe$_{2}$ crystal was identified by
single crystal x-ray diffraction (XRD). The data were collected using the
Bruker APEX2 software package\cite{Apex2} on a Bruker SMART APEX II single
crystal x-ray diffractometer with graphite-monochromated Mo K$_{\alpha }$
radiation ($\lambda $ = 0.71073 \AA ) at room temperature. Data processing
and an empirical absorption correction were also applied using APEX2
software package. The structures were solved (direct methods) and refined by
full-matrix least-squares procedures on $|F^{2}|$ using Bruker SHELXTL
program software.\cite{Sheldrick} Obtained lattice parameters of
BaFe$_{2}$OSe$_{2}$ are given in Table 1. Atomic coordinates, isotropic
displacement parameters and selected interatomic bond distances and angles
are listed in Table 2. Phase identity and purity were confirmed by powder
X-ray diffraction carried out on a Rigaku miniflex X-ray machine with Cu K$%
_{\alpha }$ radiation ($\lambda $ = 1.5418 \AA ). Structural refinements of
powder BaFe$_{2}$OSe$_{2}$\ sample was carried out by using Rietica
software. \cite{Hunter}

The average stoichiometry was determined by examination of multiple points
using an energy-dispersive X-ray spectroscopy (EDX) in a JEOL JSM-6500
scanning electron microscope. The presence of oxygen was confirmed for BaFe$%
_{2}$OSe$_{2}$, but the exact amount could not be quantified because of
experimental limitations. The average atomic ratios determined from EDX are
Ba:Fe:Se = 1.00(7):2.2(2):2.0(2), close to the ratio of stoichiometric BaFe$%
_{2}$OSe$_{2}$.

M\"{o}sbauer spectrum was taken in transmission mode with $^{57}$Co(Rh)
source at 294 K and the parameters were obtained using WinNormos software.%
\cite{Brand} Calibration of the spectrum was performed by laser and isomer
shifts were given with respect to $\alpha $-Fe.

The X-ray absorption spectra of the Fe and Se $K$-edges were taken in
transmission mode on powder samples of BaFe$_{2}$OSe$_{3}$ at the X19A
beamline of the National Synchrotron Light Source. Standard procedure was
used to extract the extended x-ray absorption fine-structure (EXAFS) from
the absorption spectrum.\cite{Prins}

Electrical transport, heat capacity, and magnetization measurements were
carried out in Quantum Design PPMS-9 and MPMS-XL5.

\section{Results and Discussion}

\begin{figure}[tbp]
\centerline{\includegraphics[scale=0.45]{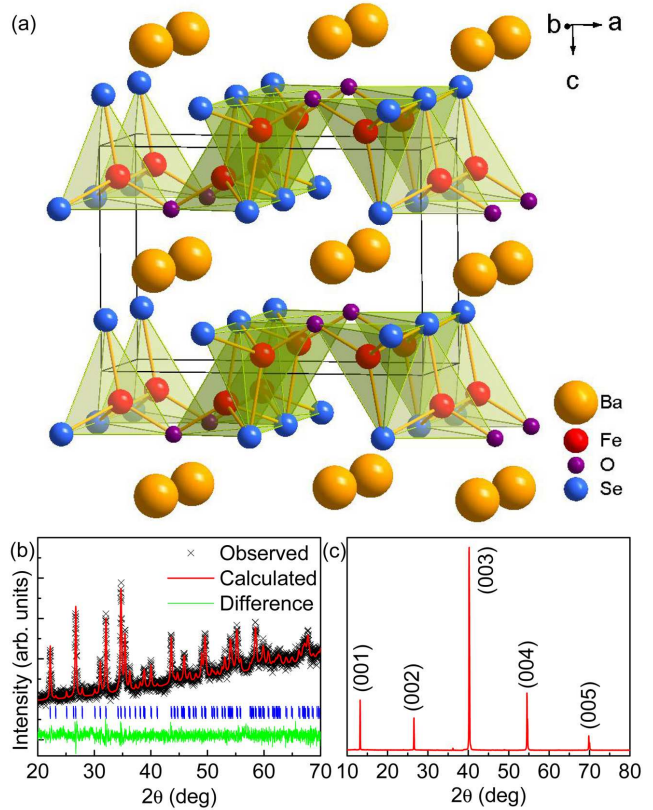}} \vspace*{-0.3cm}
\caption{(a) Crystal structure of BaFe$_{2}$OSe$_{2}$. The biggest orange,
big red, medium blue and small purple balls represent Ba, Fe, O and Se ions,
respectively. (b) and (c) Powder and single crystal XRD patterns of BaFe$%
_{2} $OSe$_{2}$.}
\end{figure}

Similar to K$_{x}$Fe$_{2-y}$Se$_{2}$ and BaFe$_{2}$Se$_{3}$,\cite{Guo}$^{,}$%
\cite{Lei HC1} the structure of BaFe$_{2}$OSe$_{2}$ is built up by stacking
the Ba cations and Fe-Se(O) layers alternatively along the c axis (Fig.
1(a)). However, within the Fe-Se(O) layers, the connection of Fe-Se(O)
tetrahedra in BaFe$_{2}$OSe$_{2}$ is distinctive different from those in K$%
_{x}$Fe$_{2-y}$Se$_{2} $ and BaFe$_{2}$Se$_{3}$. In BaFe$_{2}$OSe$_{2}$,
double chains of edge-shared Fe-Se(O) tetrahedra propagate along the Se
atoms parallel to $b$-axis. The Fe-Se(O) double chains are bridged by oxygen
along the $a$-axis (Fig. 1(a)). The bond distances of Fe-Se ($\sim $ 2.5 \AA %
) and Fe-O ($\sim $ 1.9 \AA ) are very similar to distances in compounds
where Fe is also in tetrahedral coordination (BaFe$_{2}$Se$_{3}$: $d_{Fe-Se}$
$\sim $ 2.4 \AA , Fe$_{3}$O$_{4}$: $d_{Fe-O} $ $\sim $ 1.9 \AA ). Because of
different bond distances between Fe-Se and Fe-O, Fe atoms are located at the
highly distorted tetrahedral environment when compared to other materials
with pure Fe-Se tetrahedra. This distortion is also reflected in the
significant deviation of bond angles from the value for the ideal
tetrahedron (109.5$^{\circ }$). The bond angles range up to 130.38(10)$%
^{\circ }$ in BaFe$_{2}$OSe$_{2}$, much larger than the values in BaFe$_{2}$%
Se$_{3}$ and K$_{x}$Fe$_{2-y}$Se$_{2}$.\cite{Lei HC1}$^{,}$\cite{Lei HC2}
Moreover, the different connection of Fe-Se(O) leads to the larger nearest
neighbor Fe-Fe distances ($d_{Fe-Fe}$ ($\sim $ 3.13 \AA )) when compared to
BaFe$_{2}$Se$_{3}$ and K$_{x}$Fe$_{2-y}$Se$_{2} $.\cite{Lei HC1}$^{,}$\cite%
{Lei HC2} Using obtained Fe-Se and Fe-O bond lengths as shown in Table 2,
the valence of Fe ions in BaFe$_{2}$OSe$_{2}$ can be calculated using the
bond valence sum (BVS) formalism in which each bond with a distance $d_{ij}$
contributes a valence $v_{ij}=\exp[(R_{ij}-d_{ij})/0.37]$ with $R_{ij}$ as
an empirical parameter and the total of valences of atom $i$, $V_{i}$ equals
$V_{i}=\underset{j}{\sum v_{ij}}$.\cite{Brown}$^{,}$\cite{Brese} The
calculated valence of Fe ions is +2.23, slightly larger than the apparent
oxidation state (+2) for Fe ions. Structural refinement of powder XRD
results confirms that all reflections can be indexed in the Pmmn space group
and the refined lattice parameters are $a$ = 9.862(1) \AA , $b$ = 4.138(1)
\AA , and $c$ = 6.732(1) \AA , close to the values obtained from single
crystal XRD. Only (00l) reflections were observed in the single crystal XRD
patterns of a BaFe$_{2}$OSe$_{2}$ (Fig. 1(c)), indicating that the
crystallographic c axis is perpendicular to the plane of the single crystal.

\bigskip
\begin{table}[tbp]\centering%
\caption{Crystallographic Data for BaFe$_{2}$OSe$_{2}$.}%
\begin{tabular}{cc}
\hline\hline
Chemical Formula & BaFe$_{2}$OSe$_{2}$ \\
Formula Mass (g/mol) & 422.94 \\
Crystal System & orthorhombic \\
Space Group & Pmmn (No. 59) \\
$a$ (\AA ) & 9.8518(7) \\
$b$ (\AA ) & 4.1332(3) \\
$c$ (\AA ) & 6.7188(4) \\
$V$ (\AA $^{3}$) & 273.59(3) \\
Z & 2 \\
Density (g/cm$^{3}$) & 5.134 \\
$R1$/$wR2$ ($F_{0}>4\sigma F_{0}$)/$R1$ (all data)$^{a}$ &
0.0459/0.1423/0.0476 \\
Goodness-of-Fit & 1.187 \\ \hline\hline
\multicolumn{2}{c}{$^{a}$ $R1=\Sigma |F_{0}|-|F_{c}|/\Sigma |F_{0}|$,} \\
\multicolumn{2}{c}{$wR2=[\Sigma (|F_{0}^{2}-|F_{c}^{2}|)^{2}/\Sigma
(wF_{0}^{2})^{2}]^{1/2}$.}%
\end{tabular}%
\label{1}%
\end{table}%

\begin{table*}[tbp]\centering%
\caption{Atomic Coordinates, Equivalent Isotropic Displacement Parameters,
and Selected Bond Lengths and Angles for BaFe$_{2}$OSe$_{2}$.}%
\begin{tabular}{cccccc}
\hline\hline
Atom & Wyckoff & $x$ & $y$ & $z$ & $U_{eq}$ (\AA $^{2}$) \\
Ba & 2b & 3/4 & 1/4 & 0.49047(4) & 0.00734(9) \\
Fe & 4f & 0.58580(5) & 3/4 & 0.87890(7) & 0.00915(11) \\
O & 2a & 3/4 & 3/4 & 0.7372(6) & 0.0089(5) \\
Se & 4f & 0.45734(4) & 1/4 & 0.75887(5) & 0.00739(10) \\ \hline
\multicolumn{6}{c}{Interatomic Distances (\AA )} \\
Fe-Se & 2.4706(6) & Fe-Se & 2.5540(4) & Fe-O & 1.8769(19) \\
Ba-Se & 3.3541(3) & Ba-Se & 3.4008(4) & Ba-O & 2.649(2) \\
Ba-Fe & 3.7012(4) &  &  &  &  \\ \hline
\multicolumn{6}{c}{Bond Angles ($^{\circ }$)} \\
Se-Fe-Se & 103.050(15) & Se-Fe-Se & 108.03(2) & Se-Fe-O & 130.38(10) \\
Se-Fe-O & 105.48(5) & Fe-Se-Fe & 76.950(15) & Fe-Se-Fe & 108.03(2) \\
Fe-O-Fe & 119.1(2) &  &  &  &  \\ \hline\hline
\end{tabular}%
\label{2}%
\end{table*}%

\begin{figure}[tbp]
\centerline{\includegraphics[scale=0.4]{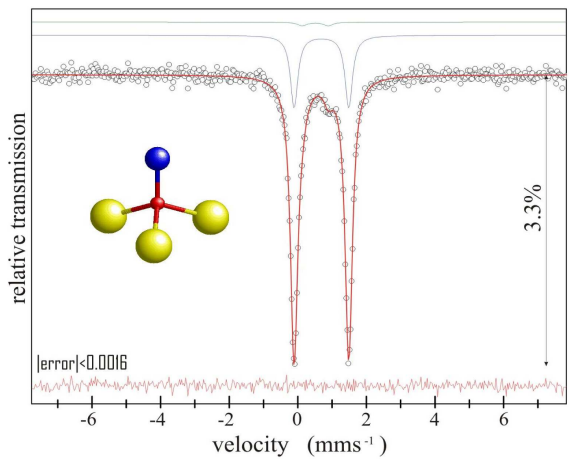}} \vspace*{-0.3cm}
\caption{M\"{o}sbauer spectrum of BaFe$_{2}$OSe$_{2}$ at T = 294 K. Inset
shows local coordination of Fe atoms in tetrahedra: central smallest ball
denote Fe, medium size ball shows oxygen wheras largest balls show selenium
atoms.}
\end{figure}

M\"{o}ssbauer fit of BaFe$_{2}$OSe$_{2}$ consists of two doublets
corresponding to the paramagnetic state (Fig. 2). Based on the goodness of
fit (0.9995) long range magnetic order is unlikely at 294 K. The first
doublet dominates the signal and takes 92 \% of the fit area with isomer
shift of 0.690(1) mm/s. First coordination sphere of Fe (inset of Fig. 2) is
tetrahedral with three selenium and one oxygen atoms. Fe$^{2+}$ and Fe$^{3+}$
in FeO$_{4}$ tetrahedra have isomer shift values (0.85-0.95) mm/s and
(0.2-0.3) mm/s respectively. Isomer shift drops with the increase of
effective charge near $^{57}$Fe. Smaller value of isomer shift in BaFe$_{2}$%
OSe$_{2}$ when compared to FeO$_{4}$ tetrahedra is consistent with larger
covalency of Fe-Se bonds. Isomer shift is in agreement with the calculate
value for Fe valence by BVS formalism. Quadrupole shift of 1.594(2) mm/s
points to large distortion of tetrahedra. Second doublet has isomer shift of
0.51(1) mm/s and 0.76(3) quadrupole splitting. These parameters are similar
to what was observed in other Fe compounds (FeSe$_{2}$ or $\alpha $-FeSe),
pointing either to presence of impurities or to intrinsic property of Fe
environment in BaFe$_{2}$OSe$_{2}$. Assuming similar recoil-free factors 8
\% of fit area corresponds to approximately similar volume (crystallographic)
phase fraction. Since such large impurity fraction would have been detected
in our X-ray experiments this might suggest that second doublet comes from
undistorted FeSe$_{4}$ tetrahedra in BaFe$_{2}$OSe$_{2}$ with no oxygen
bonds present. In other words, there might be 8\% ionic disorder existing.

\begin{figure}[tbp]
\centerline{\includegraphics[scale=0.45]{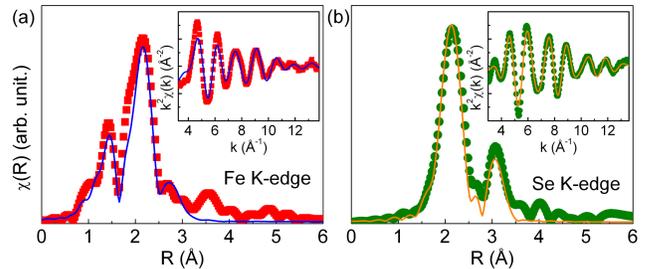}} \vspace*{-0.3cm}
\caption{FT magnitudes of the EXAFS oscillations (symbols) for Fe K-edge (a)
and Se K-edge (b). The model fits are shown as solid lines. The FTs are not
corrected for the phase shifts and represent raw experimental data. Insets
of (a) and (b) filtered EXAFS (symbols) with k-space model fits (solid
line). }
\end{figure}

The local environments of Fe and Se such as bond distances are also analyzed
by fitting the EXAFS data as show in Fig. 3. For the Fe site, the nearest
neighbors are one oxygen atom ($d_{Fe-O}=$ 1.8769(19) \AA ) and three Se
atoms (1$\times $Se(I) and 2$\times $Se(II)) with two different distances ($%
d_{Fe-Se(I)}=$ 2.4706(6) \AA\ and $d_{Fe-Se(II)}=$ 2.5540(4) \AA ).\ The
next nearest neighbors are three Fe atoms (1$\times $Fe(I) and 2$\times $%
Fe(II)) with two different distances ($d_{Fe-Fe(I)}\sim $ 3.13 \AA\ and $%
d_{Fe-Fe(II)}\sim $ 3.24 \AA ). On the other hand, for the Se site, the
nearest neighbors are three Fe atoms with bond distances $d_{Fe-Se(I)}$ and $%
d_{Fe-Se(II)}$. The next nearest neighbors are three Ba atoms (2$\times $%
Ba(I) and 1$\times $Ba(II)) with two different distances ($d_{Se-Ba(I)}=$
3.3541(3) \AA\ and $d_{Se-Ba(II)}=$ 3.4008(4) \AA ). From the joint analysis
of Fe and Se edges EXAFS data using a single bond distance for Fe-O, Fe-Se,
Fe-Fe and Se-Ba, and by fitting the k range 2-14 \AA $^{-1}$ for Fe $K$-edge
and 2-12.9 \AA $^{-1}$ for Se $K$-edge (insets of Fig. 3(a) and (b)), the
fitted average bond lengths are $d_{Fe-o}=$ 1.87(2) \AA , $d_{Fe-Se}=$\
2.500(8) \AA , $d_{Fe-Fe}=$ 3.17(4) \AA\ and $d_{Se-Ba}=$ 3.38(2) \AA ,
which are consistent with the average bond distances derived from XRD
fitting.

\begin{figure}[tbp]
\centerline{\includegraphics[scale=0.4]{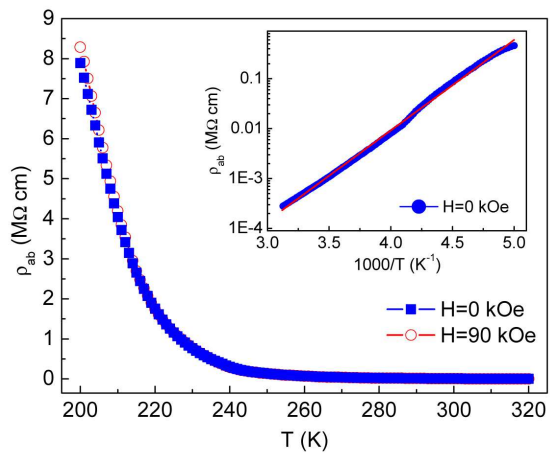}} \vspace*{-0.3cm}
\caption{Temperature dependence of the resistivity $\protect\rho _{ab}(T)$
of the BaFe$_{2}$OSe$_{2}$ crystal with $H$ = 0 (closed blue square) and 90
kOe (open red circle, H$\Vert $c). Inset shows the fitting result of $%
\protect\rho _{ab}(T)$ at zero field using thermal activation model. The red
line is the fitted curve.}
\end{figure}

As shown in Fig. 4, the resistivity $\rho _{ab}(T)$ of BaFe$_{2}$OSe$_{2}$
crystal increases rapidly with decreasing temperature, suggesting that this
compound is a semiconductor in measured temperature region. The
semiconducting behavior of $\rho _{ab}(T)$ can be fitted using the thermal
activation model $\rho _{ab}(T)=\rho _{0}\exp (E_{a}/k_{B}T)$, where $\rho
_{0}$ is a prefactor, $E_{a}$ is thermal activated energy and $k_{B}$ is the
Boltzmann's constant (inset in Fig. 4). The obtained $E_{a}$ is 0.360(2) eV\
and the room-temperature resistivity $\rho _{ab}(300K)$ is about 10 k$\Omega
\cdot $cm. Both are much larger than in BaFe$_{2}$Se$_{3}$ (0.178 eV and 17 $%
\Omega \cdot $cm).\cite{Lei HC1} The $E_{a}$ is also larger than in La$_{2}$O%
$_{2}$Fe$_{2}$OSe$_{2}$ where Fe has octahedral coordination with 4$\times $%
Se and 2$\times $O mixed anions.\cite{Zhu JX} The large $\rho _{ab}(300K)$
and $E_{a}$\ could be due to the shorter Fe-O bond in highly distorted
structure which localizes electrons and increases the band gap. Since La$%
_{2} $O$_{2}$Fe$_{2}$OSe$_{2}$ is a Mott insulator with narrow 3d electronic
bands due to strong correlation effects, it is of interest to study whether
BaFe$_{2}$OSe$_{2}$ is normal band insulator or Mott insulator via
theoretical calculations. The $\rho _{ab}(T)$ measured at $H$ = 90 kOe
indicates that there is no obvious magnetoresistance in BaFe$_{2}$OSe$_{2}$
(Fig. 4).

\begin{figure}[tbp]
\centerline{\includegraphics[scale=0.45]{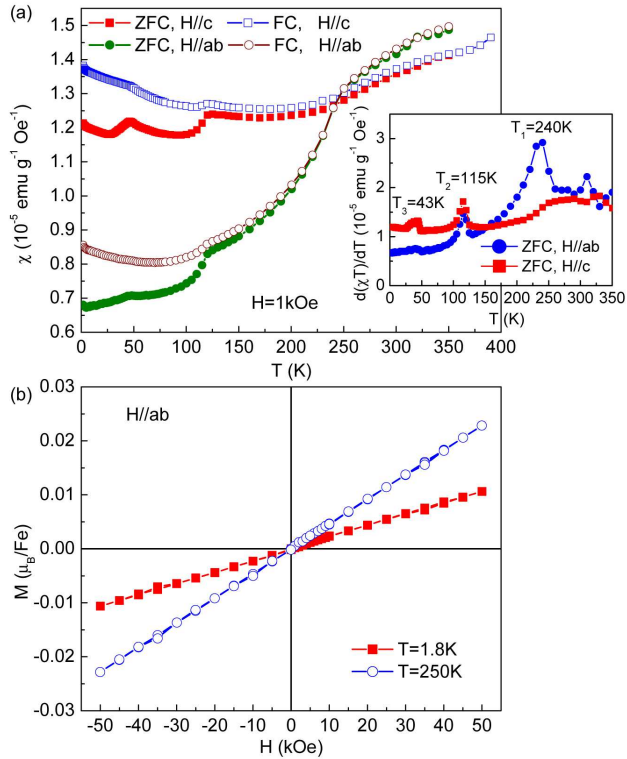}} \vspace*{-0.3cm}
\caption{(a) Temperature dependence of DC magnetic susceptibility $\protect%
\chi $(T) under ZFC and FC modes with $H$ = 1 kOe along ab-plane and c-axis.
(b) Isothermal magnetization hysteresis loops M(H) for $H\Vert ab$ at $T$ =
1.8 K and 250 K.}
\end{figure}

The Curie-Weiss temperature dependence of $\chi $(T) is not observed in BaFe$%
_{2}$OSe$_{2}$ single crystals up to 350 K (Fig. 5(a)). On the other hand, M\"{o}ssbauer fit indicate that long range magnetic order is absent at 294 K. It implies that short range magnetic interactions might be present already at the room
temperature. There is a peak in the $d(\chi T)/dT$ curve at $T_{1}$ = 240 K,%
\cite{Fisher} for H$\parallel $ and the values of susceptibility between ZFC
and FC curves are nearly the same at this temperature. This peak corresponds
to the AFM transition temperature $T_{N}$ = 240 K, confirmed by specific
heat measurements. Moreover, the decrease in $\chi _{c}(T)$ with temperature
is more significant than $\chi _{ab}(T)$. It suggests that the easy-axis of
magnetization direction could be\ in $ab$-plane. According to mean-field
theory for collinear antiferromagnets, $\chi (T)\rightarrow 0$ along the
easy-axis direction whereas it is nearly constant below $T_{N}$ for the
field perpendicular to the easy-axis direction. With further decreasing
temperature, there are two peaks in the $d(\chi T)/dT$ curve at $T_{2}$ =
115 K and $T_{3}$ = 43 K, respectively (inset of Fig. 5(a)).\ The hysteresis
between zero-field-cooling (ZFC) and field-cooling (FC) measurements implies
that they are spin-glass-like transitions. The absence of hysteresis in
isothermal $M(H)$ loops for $H\parallel ab$ (Fig. 4(b)) at $T$ = 1.8 K and
250 K indicates that there are no ferromagnetic impurities. Moreover, the
slope of $M(H)$ increases with increasing temperature consistent with the
AFM behavior observed in $\chi (T)$ curves (Fig. 5(b)).

\begin{figure}[tbp]
\centerline{\includegraphics[scale=0.4]{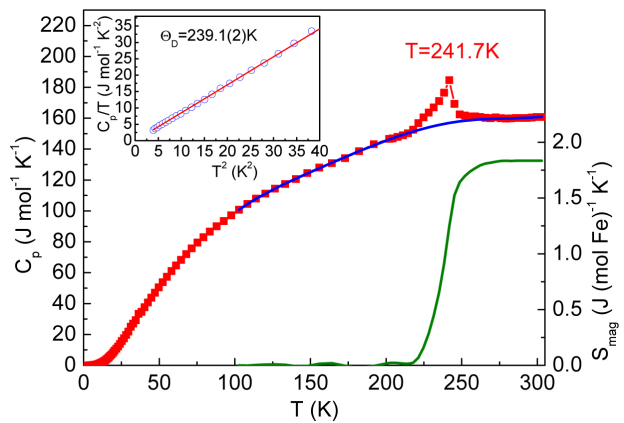}} \vspace*{-0.3cm}
\caption{Temperature dependence of specific heat for BaFe$_{2}$OSe$_{2}$
crystal. The horizontal orange line represents the classical value according
to Dulong-Petit law at high temperature with N = 6. The blue curve
represents the phonon contribution fitted by a polynomial. The right axis
and its associated green solid curve denote the magnetic entropy. Inset: the
low-temperature specific-heat data in the plot of $C_{p}/T$ \textit{vs} $%
T^{2}$. The red solid line is the fitting curves using formula $C_{p}/T=%
\protect\beta T^{2}$.}
\end{figure}

As shown in Fig. 6, the specific heat $C_{p}$ of BaFe$_{2}$OSe$_{2}$
approaches the value of 3NR at high temperature ($T$ = 300 K) assuming N =
6, where N is the atomic number in the chemical formula and R is the gas
constant (R = 8.314 J mol$^{-1}$ K$^{-1}$). It confirms the atomic numbers
of BaFe$_{2}$OSe$_{2}$ obtained from single crystal XRD fitting and
consistent with the atomic ratio measured from EDX. On the other hand, at
the low temperature, $C_{p}(T)$ curve can be fitted by a cubic term $\beta
T^{3}$ (inset of Fig. 6) and the fitted value $\beta $ is 0.852(2) mJ mol$%
^{-1}$ K$^{-4}$. According to the formula $\Theta _{D}$ = $(12\pi
^{4}NR/5\beta )^{1/3}$, Debye temperature is estimated to be $\Theta _{D}$ =
239.1(2) K. The larger $\Theta _{D}$ when compared to BaFe$_{2}$Se$_{3}$ ($%
\Theta _{D}$ = 205(1) K) can be ascribed to the smaller atomic mass of
oxygen than selenium.\cite{Lei HC1}

There is a $\lambda $-type anomaly at $T_{1}$ = 241.7 K as shown in\ Fig. 6.
The peak position is very close to the peak position ($T_{1}$ = 240 K) at $%
d\chi (T)/dT$ curve (inset of Fig. 5(a)).\ It suggests that a long-range AFM
ordering forms at this temperature. After substraction the phonon
contribution ($C_{ph}$) fitted using a polynomial for the total specific
heat, the magnetic contribution ($C_{mag}$) can be obtained. The magnetic
entropy can be calculated using the integral $S_{mag}(T)=%
\int_{0}^{T}C_{mag}/Tdt$. The derived $S_{mag}$ is $\sim $ 1.83 J/mol-K when
$T$ is up to 300 K (Fig. 6), which is much smaller than expected values R$\ln(2S+1)$ for Fe$^{2+}$ ions with low spin state ($\sim $ 20\% Rln3) and
with high spin state ($\sim $ 13.7\% Rln5) in tetrahedral crystal fields. It is even less than Rln2 for S=1/2 ($\sim $ 31.8\% Rln2), suggesting that there is short range magnetic order at higher temperature
which partially releases the magnetic entropy before any long range magnetic
transition, consistent with the M\"{o}ssbauer spectrum and magnetization measurement results. On the other hand, there is no peaks appear at $T_{3}$ = 115 K and $T_{4}$
= 43 K, confirming the glassy magnetic behaviors at those temperatures.

\section{Conclusion}

In summary, we report a discovery of a new layered iron oxychalcogenide BaFe$%
_{2}$OSe$_{2}$. The crystal structure features double chains of edge-shared
Fe-Se(O) tetrahedra that propagate along the $b$-axis. The Fe-Se(O) double
chains are bridged by oxygen along the $a$-axis. BaFe$_{2}$OSe$_{2}$\ single
crystals are magnetic semiconductors with a possible long range AFM transition at 240 K and two glassy transitions at 115 K and 43 K. The magnetic entropy up to 300 K is
much smaller than expected for Fe$^{2+}$ in tetrahedral crystal fields,
suggesting the existence of short range AFM correlation above room
temperature. Because of the interesting structure and connectivity among the
iron atoms, it is of interest to investigate the mechanism of magnetic
ground state by theoretical calculations and by neutron scattering
experiments. Moreover, doping effects on physical properties should further
reveal similarities and differences with superconducting iron selenide
compounds.

\section{Acknowledgements}

We thank Kefeng Wang for helpful discussion, and Syed Khalid for help with
XAFS measurements. Work at Brookhaven is supported by the U.S. DOE under
Contract No. DE-AC02-98CH10886 A.I.F. acknowledges support by U.S.
Department of Energy Grant DE-FG02-03ER15476. Beamline X19A at the NSLS is
supported in part by the U.S. Department of Energy Grant No
DE-FG02-05ER15688. This work has also been supported by the grant No. 171001
from the Serbian Ministry of Education and Science.

\end{document}